\documentclass[twocolumn,nopacs,floatfix,amsmath,nofootinbib,amssymb,preprintnumbers]{revtex4-2}
\usepackage{graphicx,color,dcolumn,booktabs,bm}
\usepackage{txfonts}
\usepackage{xcolor}
\usepackage[colorlinks,
            citecolor=blue,
            anchorcolor=red,
            menucolor=red,
            linkcolor=red,
            filecolor=red,
            runcolor=red,
            urlcolor=blue,
            frenchlinks=red]{hyperref}
\usepackage{float}
\usepackage{braket}
\usepackage{subfigure}
\begin{document}
\preprint{
	\vbox{
		\hbox{ADP-25-25/T1287}
}}
\title{Structure of the $\Omega^{-}(2012)$ with Hamiltonian Effective Field Theory}
\author{Fang-Chao Han$^{1,2,3}$}
\author{Zhan-Wei Liu$^{1,2,3}$}\email{liuzhanwei@lzu.edu.cn}
\author{Derek~B.~Leinweber$^4$}\email{derek.leinweber@adelaide.edu.au}
\author{Anthony W. Thomas$^4$}\email{anthony.thomas@adelaide.edu.au}
\affiliation{
$^1$School of Physical Science and Technology, Lanzhou University, Lanzhou 730000, China\\
$^2$Research Center for Hadron and CSR Physics, Lanzhou University and Institute of Modern Physics of CAS, Lanzhou 730000, China\\
$^3$Lanzhou Center for Theoretical Physics, MoE Frontiers Science Center for Rare Isotopes, Key Laboratory of Quantum Theory and Applications of MoE, Key Laboratory of Theoretical Physics of Gansu Province, Gansu Provincial Research Center for Basic Disciplines of Quantum Physics, Lanzhou University, Lanzhou 730000, China\\
$^4$CSSM,
  Department of Physics, University of Adelaide, South
  Australia 5005, Australia
}

\begin{abstract}
We investigate the internal structure of the $\Omega(2012)^-$ by analyzing lattice QCD simulation and experimental data within Hamiltonian effective field theory, considering both $J^P = 1/2^-$ and $3/2^-$ assignments. The couplings to the dominant decay channel $\Xi \bar{K}$ and the near-threshold channel $\Xi(1530) \bar{K}$ are determined through the quark-pair-creation model. By studying the lattice QCD spectra in these two spin-parity scenarios, we extract the masses and widths of the resonances. We notice that the $J^P = 3/2^-$ resonance is consistent with the observed $\Omega(2012)^-$ while the recently reported $\Omega(2109)^-$ may be a $J^P = 1/2^-$ $\Omega$.
\end{abstract}
\maketitle
\section{introduction}\label{sec1}
In 2018, the Belle Collaboration first observed the $\Omega(2012)^{-}$ in its decay channels $\Xi^0K^-$ and $\Xi^-K_s^0$ using the $e^+e^-$-collision data near the resonances $\Upsilon(1S)$, $\Upsilon(2S)$ and $\Upsilon(3S)$~\cite{Belle:2018mqs}, and further studied the three-body decay $\Omega(2012)^{-} \rightarrow \bar{K}\Xi(1530) \rightarrow \bar{K}\pi\Xi$~\cite{Belle:2019zco,Belle:2022mrg}. In 2024, evidence for the $\Omega(2012)^-$ was reported by the BESIII Collaboration in the process $e^+e^- \rightarrow \Omega(2012)^- \bar{\Omega}^+ + c.c.$ with a significance of 3.5$\sigma$~\cite{BESIII:2024eqk}. Notably, in the same experiment, a new resonance $\Omega(2109)$, was also observed with a significance of 4.1$\sigma$, providing evidence for another excited $\Omega$ hyperon. Most recently, the ALICE Collaboration reported a signal consistent with the $\Omega(2012)^{-}$ with a significance of 15 $\sigma$ in $pp$ collisions at $\sqrt{s} = 13~\text{TeV}$~\cite{ALICE:2025atb}. The $\Omega(2012)^{-}$ is classified by the Particle Data Group (PDG) as a three-star excited state of the $\Omega^{-}$ with the mass and width being \cite{ParticleDataGroup:2024cfk}
\begin{eqnarray}
M_{\Omega(2012)^{-}}&=&2012.4\pm0.7\pm0.6\textrm{ MeV},\quad\\  
\Gamma_{\Omega(2012)^{-}} &=& 6.4_{-2.0}^{+2.5}\pm1.6\textrm{ MeV}. \nonumber
\end{eqnarray}
The $\Omega(2012)^{-}$ is the first excitation in the $\Omega^{-}$ spectrum. In the naive quark model $\Omega^{-}$ is made of three identical strange quarks similar to $\Delta^{++}$ or $\Delta^{-}$. However, the first excited state $\Delta(1600)$ shares the same spin-parity as the ground-state $\Delta(1232)$ while in the $\Omega$ sector the first excited state $\Omega(2012)^-$ is proposed to have $J^P = 3/2^-$ or $1/2^-$, which is different from that of the ground-state $\Omega(1672)^-$. This reflects not only a parity flip but also a possible change in spin, suggesting a fundamentally different excitation mechanism. Investigating the strange quark system is a crucial step toward unveiling the nature of the strong interaction~\cite{Crede:2024hur,Hyodo:2020czb}.

In this context, the most pressing question concerns the uncertain nature of the $\Omega(2012)^-$, in particular whether it exhibits a quark-model-like or molecular structure, as well as its spin-parity assignment. Before the discovery of the $\Omega(2012)^{-}$, many models had already been applied to the excited states of $\Omega^{-}$ with  $J^P = 1/2^-$ or $3/2^-$~\cite{An:2013zoa,An:2014lga,Capstick:1986ter,Chao:1980em,Faustov:2015eba,Kalman:1982ut,Liu:2007yi,Loring:2001ky,Pervin:2007wa,Oh:2007cr,Engel:2013ig,Edwards:2012fx}.  The three-quark structure has been adopted to explain the $\Omega(2012)^{-}$~\cite{Xiao:2018pwe,Aliev:2018syi,Aliev:2018yjo,Wang:2018hmi,Polyakov:2018mow,Liu:2019wdr,Liu:2020yen,Arifi:2022ntc,Zhong:2022cjx,Wang:2022zja,Su:2024lzy,Aliev:2023erd,Lu:2022puv,Oudichhya:2022off,Menapara:2022iuh,Luo:2025cqs}, including the chiral quark model~\cite{Xiao:2018pwe,Zhong:2022cjx}, QCD sum rule~\cite{Aliev:2018syi,Aliev:2018yjo,Su:2024lzy}, quark-pair-creation model~\cite{Wang:2018hmi}, and so on. Moreover, the mass of the $\Omega(2012)^-$ lies only 16.5 MeV below the $\Xi(1530)\bar{K}$ threshold, which has led to interpretations of this state as a possible $\Xi(1530)\bar{K}$ molecular configuration or a dynamically generated resonance~\cite{Valderrama:2018bmv,Lin:2018nqd,Pavao:2018xub,Huang:2018wth,Gutsche:2019eoh,Lu:2020ste,Lin:2019tex,Ikeno:2020vqv,Ikeno:2022jpe,Ikeno:2023wyh,Xie:2024wbd,Song:2024ejc,Hu:2022pae,Zeng:2020och}.

Lattice QCD starts from the first principles of QCD, which is expected to provide useful insights into the internal structure of $\Omega(2012)^{-}$. In earlier studies, the Bern-Graz-Regensburg collaboration~\cite{Engel:2013ig} and hadron spectrum collaboration ~\cite{Edwards:2012fx} investigated the negative-parity $1P$ baryons $\Omega$ with spin-$1/2$ and spin-$3/2$ on the lattice. The former employed two mass-identical light quarks with the chirally improved fermion action, while the latter used an anisotropic clover action to generate gauge configurations. These results were all obtained on a smaller spatial lattice volume with $L \simeq 1.98~\text{fm}$. The $\Omega$ baryon masses were also studied by the coordinated lattice simulations collaboration, where the $N_f = 2 + 1$ gauge configurations were generated along trajectories maintaining an approximately constant trace of the bare quark mass matrix \cite{Hudspith:2024kzk}. By describing the positive-parity $\Omega$ masses with the N${}^3$LO SU(3) chiral perturbation theory expressions, the lattice scales were thus determined and then used to obtain the negative-parity masses. 

Recently, lattice QCD simulations for the $\Omega^{-}$ baryon spectrum were released by employing smeared three-quark operators on the PACS-CS configurations~\cite{Hockley:2024aym} and, notably, the nodes of corresponding radial wave functions were also identified. However, the shifts in the finite-volume energy levels caused by nearby scattering states were not analyzed and thus the widths of the resonances were not extracted. We apply Hamiltonian effective field theory (HEFT) on these results with spatial lattice volume $L \simeq 2.9~\text{fm}$ to further study the odd-parity $\Omega^{-}$ in this work.

HEFT describes resonance positions, decay widths, scattering phase shifts, and inelastic effects in the infinite volume, meanwhile it is also applicable to analyze hadron excitations from the lattice QCD simulations in the finite volume~\cite{Hall:2013qba,Hall:2014gqa,Hall:2014uca,Liu:2015ktc,Liu:2016uzk,Liu:2016wxq,Liu:2023xvy,Wu:2017qve,Abell:2021awi,Abell:2023qgj,Leinweber:2024psf,Abell:2023nex,Zhuge:2024iuw,Hockley:2024ipz,Abell:2023nex}. In particular, examining the eigenvectors of the Hamiltonian in finite volume helps probe the internal structure of coupled-channel systems. HEFT has been successfully used to study the nucleon excited states, the $\Delta$ resonance, the $\Lambda(1405)$ and $\Lambda(1670)$, and so on.

This paper is organized as follows. In the Section on framework, we outline
the HEFT framework as applied to the negative parity $\Omega(2012)^{-}$ hyperons. In the following Section the corresponding numerical results
and discussion are presented. Finally, a summary and concluding remarks are provided.
\section{framework}\label{sec2}
The spin-parity $J^P$ of $\Omega(2012)^{-}$ could be either $1/2^-$ or $3/2^-$~\cite{Xiao:2018pwe}, and we will examine both these odd-parity systems. This section introduces the framework for the relevant interaction, the T matrix at infinite volume, and matrix Hamiltonian at finite volume, from which one can obtain the masses and widths related to the resonances. 
\subsection{Interaction}\label{sec:TotalH}
The Hamiltonian reads
\begin{equation}
\label{Hamiltonian}
H=H_0+H_{int} \, ,
\end{equation}
where the kinetic-energy Hamiltonian $H_0$ is written as
\begin{align}\label{eq:Kinetic-energy hamiltonian}
H_{0}  =&\left|\Omega_{0}\right\rangle m_{\Omega}^{0}\left\langle \Omega_{0}\right|\notag\\ &+\sum_{\alpha}\int d^3\vec{k} \left|\alpha(\vec{k})\right\rangle\left[\omega_{\alpha_{M}}(k)+\omega_{\alpha_{B}}(k)\right]\left\langle\alpha(\vec{k})\right|.
\end{align}
The $m_{\Omega}^{0}$ can be interpreted as a “bare baryon” composed of a three-quark core, which is not dressed by coupling to meson-baryon channels. Here $\left|\alpha\right\rangle=\left|\Xi\bar{K}\right\rangle\,\text{or}\,\left|\Xi(1530)\bar{K}\right\rangle$, and 
\begin{equation}\label{eq:Kinetic-energy}
\omega_{\alpha_{B(M)}}(k)=\sqrt{m_{\alpha_{B(M)}}^{2}+k^{2}}
\end{equation}
is the energy of the particle. The subscripts $\alpha_B$ and $\alpha_M$ represent the baryon and meson separately in channel $\alpha$.

The interaction Hamiltonian of this system is
\begin{align}\label{eq:Interaction hamiltonian}
    H_{int} = &\sum_{\alpha} \int d^3\vec{k} \{
    |\alpha(\vec{k})\rangle g^{J}_{\alpha, \Omega_0}(\vec{k}) \langle \Omega_0|\notag\\
    &+ |\Omega_0\rangle g_{\alpha, \Omega_0}^{J \dagger}(\vec{k}) \langle \alpha(\vec{k})| 
    \},
\end{align}
where the coupling $g^{J}_{\alpha, \Omega_0}(\vec{k})$ can be determined with the phenomenological models. 

In this work we use the quark-pair-creation model to constrain these couplings with which the pole and the lattice QCD results of the $\Omega(2012)$ will be then studied. This approach has been successfully applied to the analysis for the $D^{*}_{s0}(2317)$, $D^{*}_{s1}(2460)$, and their bottom analogs in both infinite and finite volumes~\cite{Yang:2021tvc,Yang:2022vdb}. 

In the quark-pair-creation model, the operator $\hat{\mathcal{T}}$ is introduced to describe that a pair of $q\bar q$ is pulled out from the vacuum~\cite{Micu:1968mk,Carlitz:1970xb,LeYaouanc:1973ldf,LeYaouanc:1974cvx,LeYaouanc:1977gm,LeYaouanc:1977fsz,LeYaouanc:1972vsx,Morel:2002vk,Ortega:2016mms}
\begin{eqnarray}
    \hat{\mathcal{T}}&=&-3\gamma\sum_{m}\braket{1m;1-m|00}\int d^3\vec{k}_4d^3\vec{k}_5\delta^3(\vec{k}_4+\vec{k}_5)\notag\\ &&\times\mathcal{Y}_1^{m}(\frac{\vec{k}_4-\vec{k}_5}{2})\chi^{45}_{1,-m}\varphi_0^{45}\omega_0^{45}b_{4}^{\dagger}(\vec{k}_4)d_{5}^\dagger(\vec{k}_5).
\end{eqnarray}
Here, $\varphi_0^{45}$ and $\omega_0^{45}$ are the
color and flavor wave functions of the $q_4\bar{q}_5$ pair created by operator $b_{4}^{\dagger}(\vec{k}_4)d_{5}^\dagger(\vec{k}_5)$ from the vacuum. $\chi^{45}_{1,-m}$ represents the spin triplet state and the solid harmonic polynomial $\mathcal{Y}_1^{m}(\vec{k})\equiv|\vec{k}|Y_1^m(\theta_k,\phi_k)$ reflects the momentum-space distribution of the $P$-wave quark pair. The helicity amplitudes can be obtained with the wave functions in quark model
\begin{equation}
    \mathcal{M}^{M_{\Omega_0}M_{\Xi}M_{\bar K}}(\vec k) =  \langle \Xi\bar K\,|\hat{\mathcal{T}}|\,\Omega_0\rangle,
\end{equation}
and the details can be found in Ref.~\cite{Wang:2018hmi} where $\gamma=6.95$ is used. With the same approach and the wave function of $\Xi(1530)$ in Ref.~\cite{Wang:2018hmi}, we can also calculate
\begin{equation}
    \mathcal{M}^{M_{\Omega_0}M_{\Xi(1530)}M_{\bar K}}(\vec k) =  \langle \Xi(1530)\bar K\,|\hat{\mathcal{T}}|\,\Omega_0\rangle.
\end{equation}

The partial wave amplitude $\mathcal{M}^{JL}(k)$ is also used in this work
\begin{eqnarray}
\label{eqn:Partial wave amplitudes}
\mathcal{M}^{JL}(k)&=\frac{\sqrt{4\pi(2L+1)}}{2J+1}\underset{M_{B},M_{C}}{\sum}\langle L0;J_{BC}M|JM\rangle
\nonumber \\&\times\langle J_{B}M_{B};J_{C}M_{C}|J_{BC}M\rangle\mathcal{M}^{MM_{B}M_{C}}(k \hat z),
\end{eqnarray}
where $J$/$J_B$/$J_C$ and $M$/$M_B$/$M_C$ represent the total angular momenta and their third components of $\Omega_0$/$B$/$C$, correspondingly. $L$ is the orbital angular momentum between the final states $B$ and $C$ and $J_{BC}$ is their total spin. Additionally, the exponential form factor with the cutoﬀ $\Lambda=1$ GeV is used to truncate the hard vertices \cite{Morel:2002vk,Ortega:2016mms}, and finally
\begin{equation}
\label{eq:gfin}
g^{J}_{\alpha, \Omega_0}(k) = 
\mathcal{M}^{JL}(k) e^{-\frac{k^2}{2\Lambda^{2}}}.
\end{equation}
%
\subsection{T Matrix at infinite volume}\label{subsec22}
The T matrix for two particle scattering can be obtained by solving a three-dimensional reduction of the coupled-channel Bethe-Salpeter equations
\begin{eqnarray}
    T_{\alpha,\beta}(k, k'; E)& = &V_{\alpha,\beta}(k, k'; E) \\&+& \sum_{\lambda} \int q^2 dq \notag
\frac{V_{\alpha,\lambda}(k, q; E)}{E - \omega_{\lambda}(q) + i\epsilon} T_{\lambda,\beta}(q, k'; E),
\end{eqnarray}
where $\omega_{\lambda}(k)$ is the center-of-mass energy of channel $\lambda$
\begin{equation}
    \omega_{\lambda}(k)=\sqrt{m_{\lambda_B}^{2}+k^2}+\sqrt{m_{\lambda_M}^{2}+k^2}.
\end{equation}
The coupled-channel potential can be calculated from the interaction Hamiltonian
\begin{equation}
    V_{\alpha,\beta}(k, k';E) = g_{\alpha,\Omega_0}^{J\dagger }(k) \frac{1}{E - m_{\Omega}^{0}} g^J_{\beta,\Omega_0}(k').
    \label{9}
\end{equation}

The pole position of $\Omega(2012)$ can be determined by locating
the pole of the T matrix. The pole is located on the second Riemann sheet. More specifically, we replace the integration variable $q$ with $q\times\exp[-i\theta]$ only for the $\Xi \bar{K}$ channel. Due to the form factor $\exp[-\frac{q^2}{2\Lambda^{2}}]$, we maintain $0 \ll \theta < \pi / 4$ to prevent exponential divergence.
\subsection{Matrix Hamiltonian at finite volume}\label{subsec23}
In the finite volume, the matrix Hamiltonian $\mathcal{H}$ is also composed of a free part and an interaction part. They can be obtained by discretizing the Hamiltonians in Sec. \ref{sec:TotalH}. 

In the system with $3/2^-$, the free $\mathcal{H}_{0}$ is
\begin{equation}
\label{eqn:H0}
 \mathcal{H}_0  = \text{diag}\,\{ m_{\Omega}^{0}\,,\,\omega_{\Xi(1530)\bar{K}}(k_0)\,,\,\omega_{\Xi\bar{K}}(k_1)\,,\,\omega_{\Xi(1530)\bar{K}}(k_1)\,,\,\cdots \}
\end{equation}
where the momenta $k_n$ can only take discrete values corresponding to a box of length $L$, $k_n^2 = \left(\frac{2\pi}{L}\right)^2 n$ with $n\equiv n_x^2 + n_y^2 + n_z^2$ and $n_i$ being the integer.

The interacting $3/2^-$ Hamiltonian $\mathcal{H}_{int}$ can be written as
\begin{equation}
\label{HI}
\scalebox{0.9}{$
\begin{pmatrix}
0 & \tilde{g}^{\frac{3}{2}\dagger}_{\Xi(1530)\bar{K},\Omega_0}(k_0) & \tilde{g}^{\frac{3}{2}\dagger}_{\Xi\bar{K},\Omega_0}(k_1) & \tilde{g}^{\frac{3}{2}\dagger}_{\Xi(1530)\bar{K},\Omega_0}(k_1) & \cdots \\
\tilde{g}^{\frac{3}{2}}_{\Xi(1530)\bar{K},\Omega_0}(k_0) &  &  &  & \cdots \\
\tilde{g}^{\frac{3}{2}}_{\Xi\bar{K},\Omega_0}(k_1) &  &0  &  & \cdots \\
\tilde{g}^{\frac{3}{2}}_{\Xi(1530)\bar{K},\Omega_0}(k_1) &  &  &  & \cdots \\
\vdots & \vdots & \vdots & \vdots & \ddots
\end{pmatrix}
$}
\end{equation}

where
\begin{equation}
\label{eqn:gfin}
\tilde{g}^J_{\alpha,\Omega_0}(k_n) = 
\sqrt{\frac{C_3(n)}{4\pi}}\left(\frac{2\pi}{L}\right)^{3/2}
g^J_{\alpha, \Omega_0}(k_n). 
\end{equation}
The $g^J_{\alpha, \Omega_0}(k)$ can be derived from Eq.~(\ref{eq:gfin}) and $C_3(n)$ denotes the number of ways in which the sum of the squares of three integers equals $n$.

We emphasize that the sum over $n$ starts from $n=0$ for S wave whereas from $n=1$ for higher partial waves since the higher-partial-wave interaction vanishes for zero momentum. For $J^P = 3/2^-$ $\Omega$, the dominant decay channel, $\Xi \bar{K}$, proceeds via $D$ wave, while the near-threshold channel $\Xi(1530) \bar{K}$ decays via $S$ wave. In contrast, the partial-wave assignments are reversed for the $J^P = 1/2^-$ system, with the  corresponding adjustments to
Eqs.~(\ref{eqn:H0}), (\ref{HI}).

One can obtain the mass spectrum in the finite volume by solving the eigenvalues of $\mathcal{H}=\mathcal{H}_{0}+\mathcal{H}_{int}$. To study the spectra of odd-parity $\Omega$ at the larger pion masses, we need to know the masses of the basic hadrons. For the mass of $m_{\Xi}(m_{\pi}^{2})$ and $m_{\bar{K}}(m_{\pi}^{2})$, we employ a smooth interpolation of the corresponding lattice QCD results \cite{Hall:2014uca,Menadue:2011pd,PACS-CS:2008bkb}. We assign a positive slope $\kappa_{\Xi(1530)}$ to $m_{\Xi(1530)}(m_{\pi}^2)$
\begin{equation}
m_{\Xi(1530)}(m_\pi^2)=m_{\Xi(1530)}|_{\mathrm{phys}}+\kappa_{\Xi(1530)}\left(m_\pi^2-m_\pi^2|_{\mathrm{phys}}\right).
\label{15}
\end{equation}
We fix $m_{\Xi(1530)}|_{\mathrm{phys}} = 1.533$ GeV from the PDG \cite{ParticleDataGroup:2024cfk}. Due to the lack of $\Xi(1530)$ lattice QCD data, we relate its slope to that of $\Delta(1232)$ since they have same spin parity. The $\Delta(1232)$ slope $\kappa_{\Delta(1232)} = 0.972\,\text{GeV}^{-1}$ was provided in Ref. \cite{Hockley:2024ipz}. $\kappa_{\Xi(1530)}=\kappa_{\Delta(1232)}/3= 0.324\,\text{GeV}^{-1}$ is approximately used because the $\Xi(1530)$ contains one light quark whereas the $\Delta(1232)$ has three light quarks in the naive quark model. Similarly for the bare $\Omega$ baryon, we still introduce
\begin{equation}\label{eq18}
m_{\Omega}^{0}(m_\pi^2)=m_{\Omega}^{0}|_{\mathrm{phys}}+\alpha\left(m_\pi^2-m_\pi^2|_{\mathrm{phys}}\right).
\end{equation}
Since the Sommer scheme \cite{Sommer:1993ce} is used to set the lattice spacing in lattice QCD data, the strange quark mass increases slightly with the light quark masses and therefore a slope parameter is necessary in any fits.  
\section{Numerical results and discussion}\label{sec3}
In this section, we study the energy spectra of $\Omega^{-}$ hyperons with $J^P = 3/2^-$ and $J^P = 1/2^-$. We consider two channels $\Xi\bar{K}$ and $\Xi(1530)\bar{K}$ and a bare baryon core to account for the three-quark configuration. We first fit the lattice QCD data in the finite volume to constrain the parameters of HEFT and then extract the pole position of the $\Omega^-$ resonance in the infinite volume. By comparing the obtained masses and widths with experiment, the preferred spin-parity of $\Omega(2012)^-$ will be analyzed. 
\subsection{The $J^P = 3/2^-$ system}
\begin{figure}[tb]
\center
\includegraphics[width=3.4in]{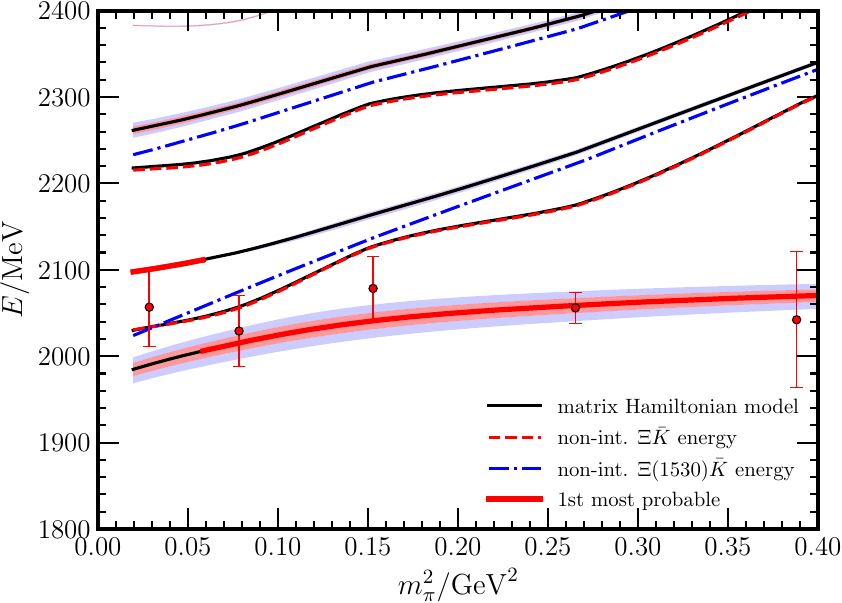}
\caption{Pion-mass dependence of the finite-volume energy for the $J^P=3/2^-$ system with $m_{\Omega}^{0}|_{\mathrm{phys}} = 2.110 \,\text{GeV}$ and the mass slope $\alpha = 0$. The solid red line indicates the largest contribution of the bare basis state in the Hamiltonian model eigenvector. The broken line denotes noninteracting meson-baryon energies. The brown and blue shaded regions represent the impacts on the finite-volume energies induced by $10\%$ and $20\%$ variations of the creation strength $\gamma$ in the ${}^3P_0$ model, respectively. The lattice results are taken from the CSSM group~\cite{Hockley:2024aym} in $2+1$ flavor QCD~\cite{PACS-CS:2008bkb}.}
\label{ben3}
\end{figure}

\begin{figure}[tb]
\centering
\subfigure[1st eigenstate]{\includegraphics[width=0.23\textwidth]{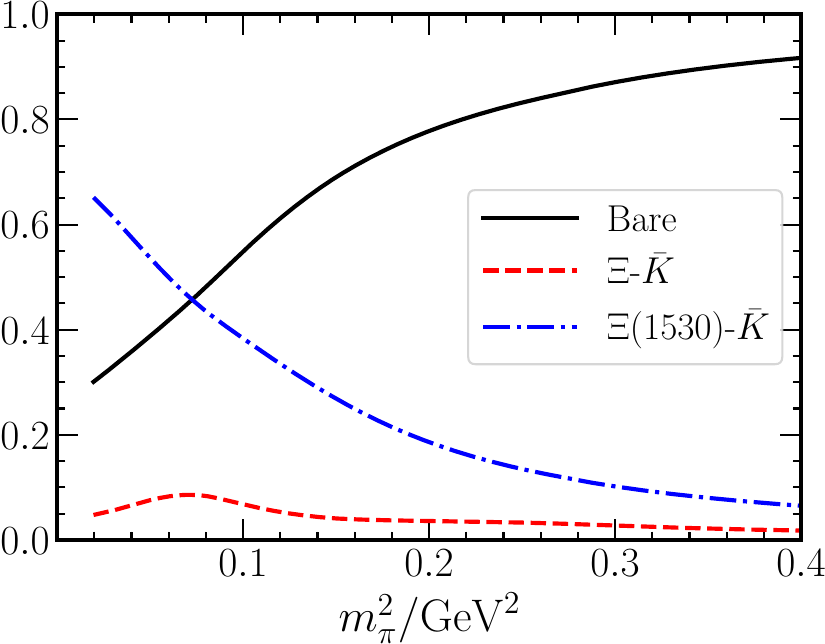}}
\subfigure[2nd eigenstate]{\includegraphics[width=0.23\textwidth]{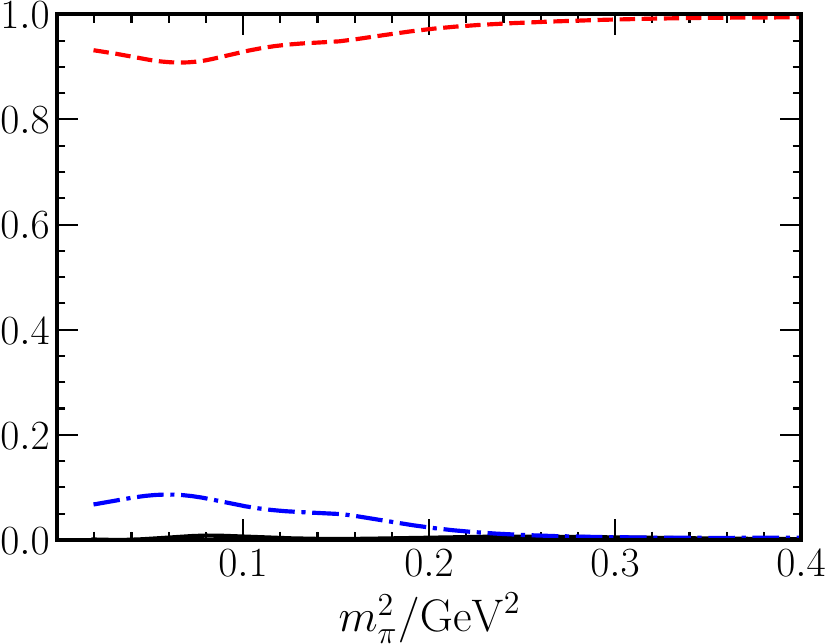}}
\subfigure[3rd eigenstate]{\includegraphics[width=0.23\textwidth]{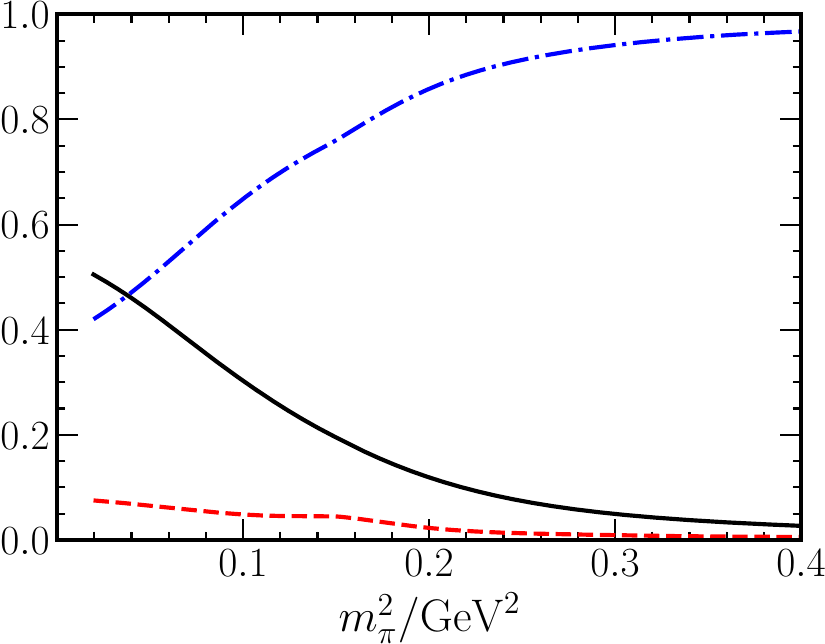}}
\subfigure[4th eigenstate]{\includegraphics[width=0.23\textwidth]{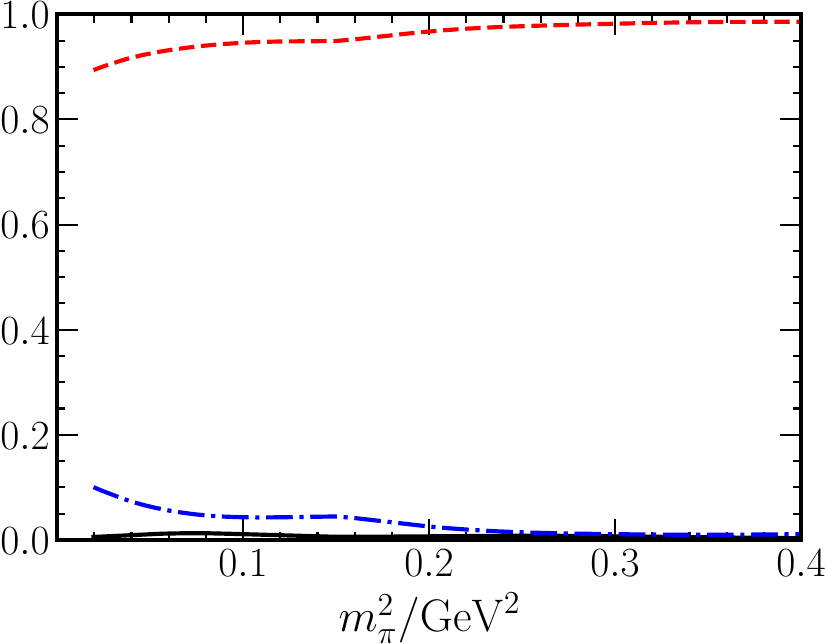}}
\caption{Pion-mass dependence of the eigenvector components with HEFT for the lowest four eigenstates of the $J^P = 3/2^-$ system.}
\label{vector3}
\end{figure}

The lattice QCD data for the masses of the $J^P = 3/2^-$ $\Omega$  are plotted as a function of $m_\pi^2$ in Fig.~\ref{ben3}. We can use HEFT to fit them well, which gives the bare mass $m_{\Omega}^{0}|_{\mathrm{phys}} = 2.110 \pm 0.018\,\text{GeV}$ and the slope $\alpha = 0.000 \pm 0.067\,\text{GeV}^{-1}$. 
The slope with the central value being zero is consistent with the fact that the bare $\Omega$ contains no light valence quarks, but the uncertainty allows for a small nontrivial slope which is to be expected in the Sommer scheme. We plot the best-fit finite-volume eigenvalues and components of the eigenstates with HEFT in Figs.~\ref{ben3} and \ref{vector3}, respectively. 

From Fig.~\ref{vector3}, for the ground state in the small mass region the largest component is $\Xi(1530)\bar{K}$, albeit with a significant bare state contribution. The bare core contribution dominates at larger pion masses.
The second eigenstate consists almost entirely of $\Xi\bar{K}$ and only a little bare core and thus it would not be easily observed on the lattice using local three-quark interpolating operators. The third eigenstate is sequentially dominated from the bare core to $\Xi(1530)\bar{K}$ as the pion mass increases. The fourth eigenstate is mainly made up of $\Xi\bar{K}$. 
 
Based on Fig.~\ref{vector3}, we highlight the low-lying states with the largest bare $\Omega$ contribution using red bold lines in Fig.~\ref{ben3}. These states should be the most likely to be observed on the lattice using local three-quark operators. One notices that all the five lattice QCD data are close to the corresponding red bold curve within one error bar. Indeed, the dominance of the bare basis state in the third finite-volume eigenstate
provides an explanation for the increased energy observed in the lattice QCD results as
the lightest quark mass is approached. Thus, HEFT is successful in explaining why the lattice QCD spectrum in this region is observed. 

We also present the non interacting energy levels in Fig.~\ref{ben3}, with the $\Xi \bar{K}$ and $\Xi(1530) \bar{K}$ channels shown as red dashed and blue dash-dotted lines, respectively. The big gap between the eigenenergies of the HEFT and the noninteracting meson-baryon energy levels reflects the presence of complicated interactions associated with resonance. With the parameters of HEFT constrained well by the lattice QCD data, we can use it to search for the resonance pole in the infinite volume.

We find a pole of the T matrix at $(2012\pm8) -(1.6\pm0.2) \,i\,\text{MeV}$, meaning that the mass is around 2012 MeV and the width is $3.2\pm0.4$ MeV. Therefore, the preferred spin parity of the $\Omega(2012)^-$ is $J^P=3/2^-$ in our approach.

These results indicate that the $\Xi(1530)\bar{K}$ channel plays a crucial role in shaping the structure and generating the real part of the $\Omega(2012)^{-}$ pole within the HEFT framework, as evidenced by both the finite-volume energy spectrum and the position of the resonance pole. Meanwhile, the $\Xi\bar{K}$ channel is responsible for providing the imaginary part of the pole, and thus contributes to the resonance width. Both channels, together with the bare core, are indispensable for achieving a consistent and realistic description of the $\Omega(2012)^{-}$ resonance.

\subsection{The $J^P = 1/2^-$ system}

\begin{figure}[tb]
\center
\includegraphics[width=3.4in]{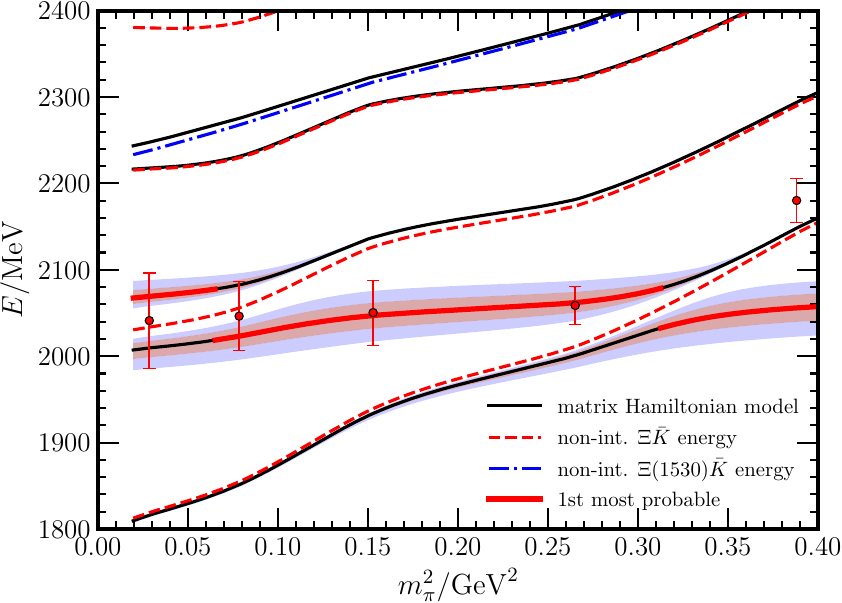}
\caption{Pion-mass dependence of the finite-volume energy for the $J^P=1/2^-$ system with $m_{\Omega}^{0}|_{\mathrm{phys}} = 2.150 \,\text{GeV}$ and the mass slope $\alpha = 0$. The solid red line indicates the largest contribution of the bare basis state within HEFT. The broken line denotes noninteracting meson-baryon energies. The brown and blue shaded regions represent the impacts on the finite-volume energies induced by $10\%$ and $20\%$ variations of the creation strength $\gamma$ in the ${}^3P_0$ model, respectively. The lattice results are taken from the CSSM group~\cite{Hockley:2024aym} in $2+1$ flavor QCD~\cite{PACS-CS:2008bkb}.}
\label{ben3-2}
\end{figure}

\begin{figure}[tb]
\centering
\subfigure[1st eigenstate]{\includegraphics[width=0.23\textwidth]{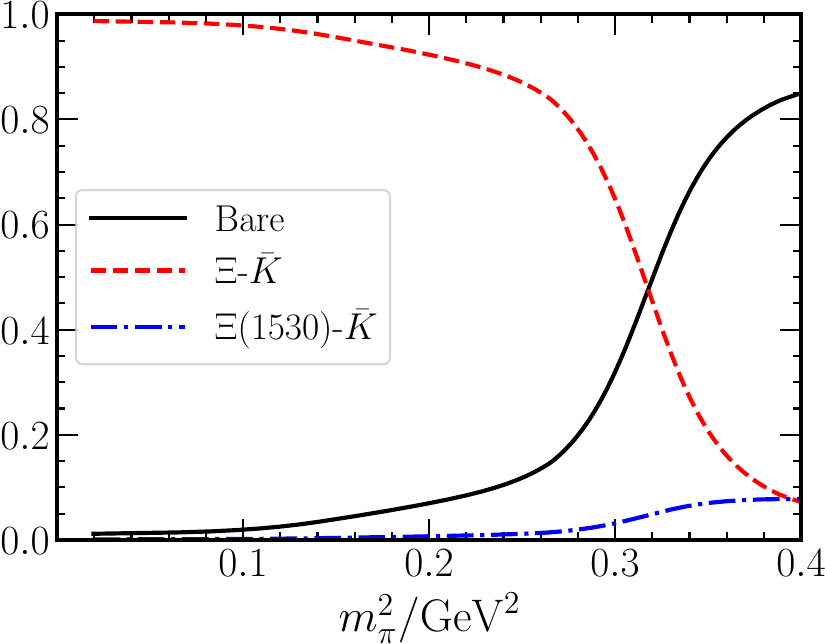}}
\subfigure[2nd eigenstate]{\includegraphics[width=0.23\textwidth]{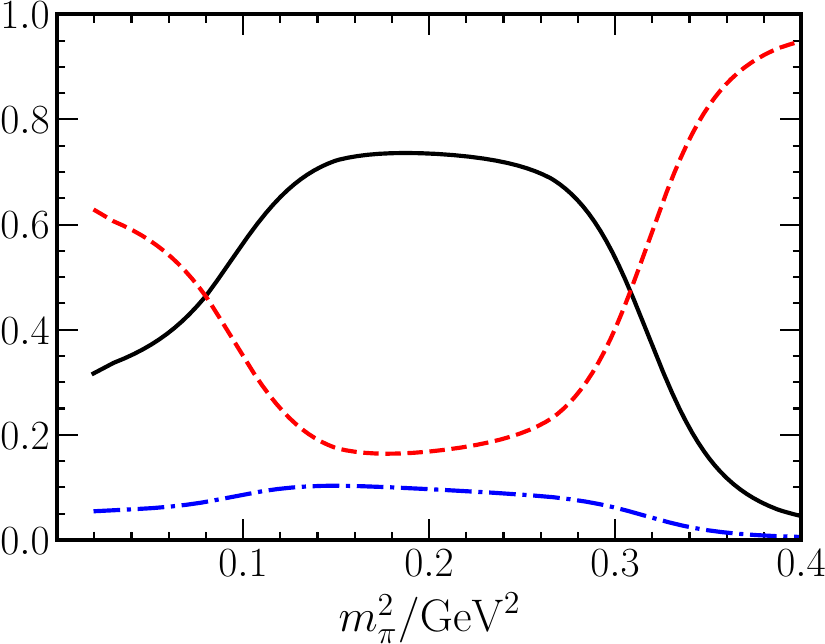}}
\subfigure[3rd eigenstate]{\includegraphics[width=0.23\textwidth]{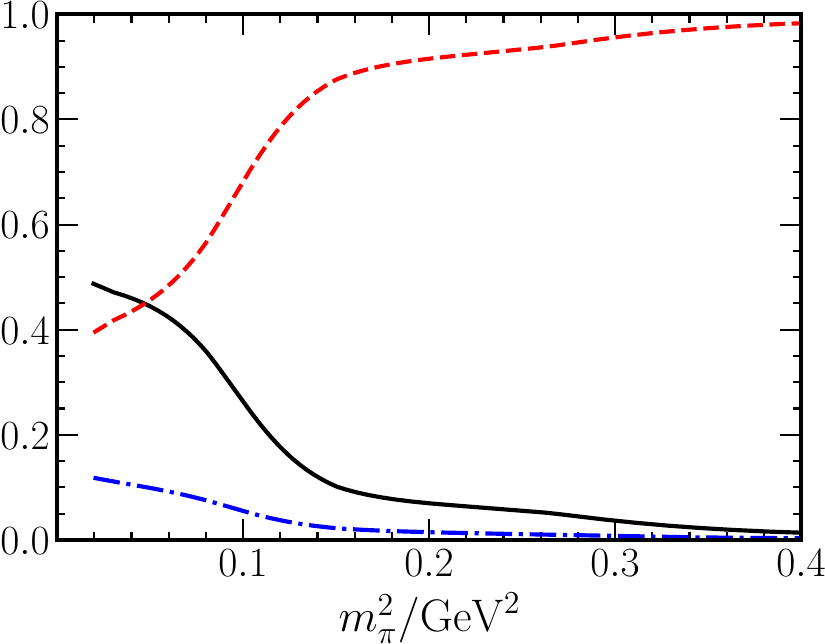}}
\subfigure[4th eigenstate]{\includegraphics[width=0.23\textwidth]{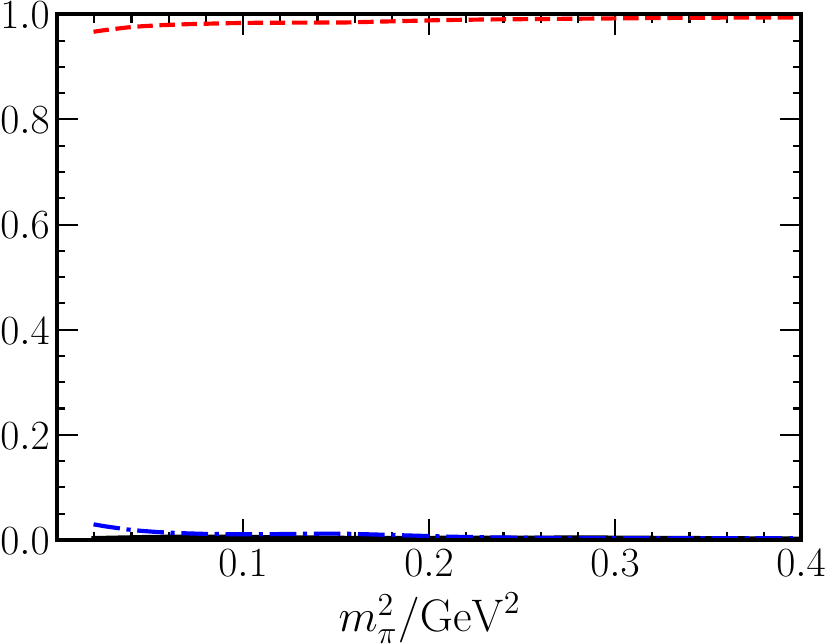}}
\caption{Pion-mass dependence of the eigenvector components with HEFT for the lowest four eigenstates of the $J^P = 1/2^-$ system.}
\label{vector3-2}
\end{figure}

Our results for the $J^P = 1/2^-$ system are presented in Figs.~\ref{ben3-2} and~\ref{vector3-2}. The lattice QCD data at small pion masses are very close to the HEFT spectrum, and the datum at the largest pion mass deviates from the second eigenenergy of HEFT by just over $1\sigma$. Figure~\ref{vector3-2} shows that the $D$-wave $\Xi(1530) \bar{K}$ channel contributes negligibly across all of the lowest four eigenstates. In general, our model can describe this system near the physical region very well.  

From Fig.~\ref{ben3-2}, the lattice QCD data typically lie near the red bold lines, with the exception of the last datum. Our HEFT analysis suggests that an early plateau associated with the second excited state was fit, and further Euclidean time evolution is required to resolve the lowest lying state.

In view of these considerations, we constrain ourselves to the region $m_{\pi}\leqslant520\,\text{MeV}$ in the fit. With the mass slope also included in the fit for the first four lattice QCD data, we obtain $m_{\Omega}^{0}|_{\mathrm{phys}} = 2.150 \pm 0.024\,\text{GeV}$ and $\alpha = 0.000 \pm 0.144\,\text{GeV}^{-1}$. This gives the pole at $(2052\pm20) - (13\pm2)\,i\,\text{MeV}$ with a slightly larger error bar for the $J^P = 1/2^-$ resonance. In this case, the width is much larger than that of the $\Omega(2012)^-$. 

However, recently the BESIII collaboration reported a new excited state, the $\Omega(2109)$, having a significance of 4.1$\sigma$, with mass $2108.5\pm5.2_{\text{stat}}\pm0.9_{\text{syst}}\,\text{MeV}$ and width $18.3\pm16.4_{\text{stat}}\pm5.7_{\text{syst}}\,\text{MeV}$ \cite{BESIII:2024eqk}. Given the larger mass and broader width, our analysis
recommends the $\Omega(2109)$ is assigned spin-parity $1/2^-$. This needs further experimental confirmation through a direct measurement of the spin and parity.
\subsection{The sensitivity analysis of creation strength $\gamma$}
The dimensionless parameter $\gamma$ characterizes the creation strength of the quark–antiquark pair in the $^{3}P_{0}$ model.  We adopt the established value of $\gamma=6.95$ reported by different groups in Refs.~\cite{Godfrey:2015dva,Godfrey:2015dia,Wang:2018hmi,Xiao:2017dly,Zhou:2023wrf}. However, there should be some uncertainty in determining this strength. Thus, we have explored the effect of changing it by 10 or 20$\%$ in order to see the sensitivity of our results. Its impacts on the finite-volume spectra are plotted as brown and  blue shades in Figs.~\ref{ben3} and \ref{ben3-2}. As can be seen in the figures, such variations of $\gamma$ induce small changes in the finite-volume energies which are less than the error bars of lattice QCD data.

The sensitivity of $\gamma$ also brings the uncertainties of poles in the infinite volume. We list the poles with the error bars
\begin{align*}
 &[2012 \pm8_{\Delta m_\Omega^0}\pm9_{10\% \gamma}(17_{20\% \gamma})] 
 \\-&[1.6\pm0.2_{\Delta m_\Omega^0}\pm0.4_{10\% \gamma}(0.8_{20\% \gamma})]i~{\rm MeV},
 &J^P=3/2^-,\\
 &[2052 \pm20_{\Delta m_\Omega^0}\pm16_{10\% \gamma}(32_{20\% \gamma})] 
 \\-&[14\pm2_{\Delta m_\Omega^0}\pm4_{10\% \gamma}(8_{20\% \gamma})]i~{\rm MeV},
 &J^P=1/2^-,
\end{align*}
where the subscripts refer to the originations of the uncertainties. One can notice that the changes of poles resulting from 10\% variation of $\gamma$ are close to those from the $\Delta m_\Omega^0$ given by the previous fit program.
%
\section{Summary}\label{summary}

In this work, we have investigated the odd-parity $\Omega$ baryons with HEFT, including the bare $\Omega$ and the channels $\Xi \bar{K}$ and $\Xi(1530) \bar{K}$. The finite-volume spectra with $J^P = 1/2^-$ and $3/2^-$ were analyzed and the pole positions subsequently extracted from the infinite-volume T matrix.

The lattice QCD simulations provide valuable information for these $\Omega^{-}$ states. HEFT describes both the lattice QCD data and the relevant resonances very well. The good description of $\Omega$ spectra indicates that the mass slope of the $\Xi(1530)$ based on quark composition is reasonable. This offers a reference point for future lattice QCD studies of $\Xi$ baryons and their excited states.
Based on the current analyses, there exist two odd-parity $\Omega$ baryon resonances with the pole positions around 2.0$\sim$2.2 GeV. The resonance with $3/2^-$ has a narrow width and the other with $1/2^-$ has a broader width. This can be understood simply as the threshold of the decay channel $\Xi(1530)\bar K$ is very close to the masses of the resonances and thus suppressed by the phasespace. In addition, the $\Xi\bar K$ channel is in $D$ wave for $3/2^-$ and $S$ wave for $1/2^-$. 
Therefore, it is expected that the $3/2^-$ state exhibits a narrower width.

The accuracy of the extracted pole position for $3/2^-$ is better than that for $1/2^-$. We find a pole of the T matrix at $(2012\pm8) -(1.6\pm0.2) \,i\,\text{MeV}$, meaning that the mass is around 2012 MeV and the width is $3.2\pm0.4$ MeV. Therefore, the preferred spin-parity of the $\Omega(2012)^-$ is $J^P=3/2^-$ in our approach.

For the $J^P = 1/2^-$ resonance, the pole extracted from the lattice QCD data is $(2052\pm20) - (13\pm2)\,i\,\text{MeV}$. Compared with the experimental mass and width, it is natural to associate $\Omega(2109)^-$ with the assignment $J^P = 1/2^-$ due to the higher mass and larger width.

 As more data is generated on the lattice by using different interpolating operators including momentum projected two-particle meson-baryon interpolating fields, the assignment of the $\Omega(2109)^-$ with $J^P = 1/2^-$ may be improved. With more precise lattice QCD and experimental data in the future, we may also incorporate additional channels such as $\Omega\eta$ and $\Xi \bar{K}^*(892)$ to further refine the description of the $\Omega(2012)^-$ and explore higher excited states. 

\section{Acknowledgment}
This work is supported by the National Natural Science Foundation of China under Grants No. 12175091, No. 12335001, No. 12247101, the ‘111 Center’ under Grant No. B20063, and the innovation project for young science and technology talents of Lanzhou city under Grant No. 2023-QN-107. This research was supported by the University of Adelaide and by the Australian Research Council through Discovery Projects No. DP210103706 (DBL) and No. DP230101791 (AWT).

\bibliography{ref}
\end{document}